**Title:**

The spatial scale dimension of speech processing in the human brain

**Authors:**

Philipp Kellmeyer[a,b,c], Roland Eric Berkemeier[d], Tonio Ball[c,e]

**Authors' affiliations:**

a Freiburg Institute for Advanced Studies (FRIAS), University of Freiburg, Albertstr. 19, D-79104 Freiburg im Breisgau, Germany

b Department of Neurosurgery, University Medical Center Freiburg

c Cluster of Excellence BrainLinks-BrainTools, University of Freiburg, Georges-Köhler-Allee 80, D-79110 Freiburg im Breisgau, Germany

d Faculty of Biology, University of Freiburg, Schänzlestraße 1, D-79104 Freiburg im Breisgau, Germany

e Neuromedical AI Lab, Department of Neurosurgery, Medical Center - University of Freiburg, Engelbergerstr. 21, D-79106 Freiburg im Breisgau, Germany

**Abstract**

In the past three decades, neuroimaging has provided important insights into structure-function relationships in the human brain. Recently, however, the methods for analyzing functional magnetic resonance imaging (fMRI) data have come under scrutiny, with studies questioning cross-software comparability, the validity of statistical inference and interpretation, and the influence of the spatial filter size on neuroimaging analyses.

As most fMRI studies only use a single filter for analysis, much information on the size and shape of the BOLD signal in Gaussian scale space remains hidden and constrains the interpretation of fMRI studies. To investigate the influence of the spatial observation scale on fMRI analysis, we use a spatial multiscale analysis with a range of Gaussian filters from 1–20 mm (full width at half maximum) to analyze fMRI data from a speech repetition paradigm in 25 healthy subjects.

We show that analyzing the fMRI data over a range of Gaussian filter kernels reveals substantial variability in the neuroanatomical localization and the average signal strength and size of suprathreshold clusters depending on the filter size. We also demonstrate how small spatial filters bias the results towards subcortical and cerebellar clusters. Furthermore, we describe substantially different scale-dependent cluster size dynamics between cortical and cerebellar clusters.

We discuss how spatial multiscale analysis may substantially improve the interpretation of fMRI data. We propose to further develop a spatial multiscale analysis to fully explore the deep structure of the BOLD signal in Gaussian scale space.

**Significance statement**

Neuroimaging data is usually analyzed with a single filter size for data pre-processing. This spatial monoscale analysis biases the results and may lead to false inferences about the data. In this study, we demonstrate how spatial multiscale filtering may improve the analysis and interpretation of fMRI data by exploring the spatial scale dimension of speech processing in the human brain. We show how the variability of neuroanatomical localization and thus the interpretation of speech-related brain activity depends on the spatial observation scale. By introducing the spatial observation scale as a new dimension in the analysis of neuroimaging data, this research opens new avenues to study cognitive processes, such as language, in the human brain.

**Introduction**

In the last two decades, neuroimaging research has provided important insights into structure-function relationships in the human brain. An examination of databases of scholarly papers reveals that in the year 2015 alone, more than 29,000 studies with fMRI were published in peer-reviewed academic journals.

Recently, however, the methods and software used for the analysis of fMRI data have come under increased scrutiny. From these investigations, several lines of enquiry emerge: the cross-software comparability (1–3), the validity of statistical inference (4–6) and interpretation (7), and the influence of the spatial observation scale on the anatomical assignment of activation peaks (8–10). Here, we focus on the influence of the spatial observation scale on the analysis of speech processing using a simple language task in fMRI.

Common justifications for spatial filtering of the blood-oxygen-level dependent (BOLD) signal are (a) to ensure that the Gaussian random field assumptions underlying the typical fMRI statistics are fulfilled, (b) to reduce noise in the signal, (c) to increase the chances that areas that are functionally homologous yet anatomically separate between individuals still overlap in group analyses, and (d) to ensure an optimal signal-to-noise ratio (SNR) by matching the filter width to the expected signal width, thus applying the matched filter theorem (11) (see **Fig. 1**).

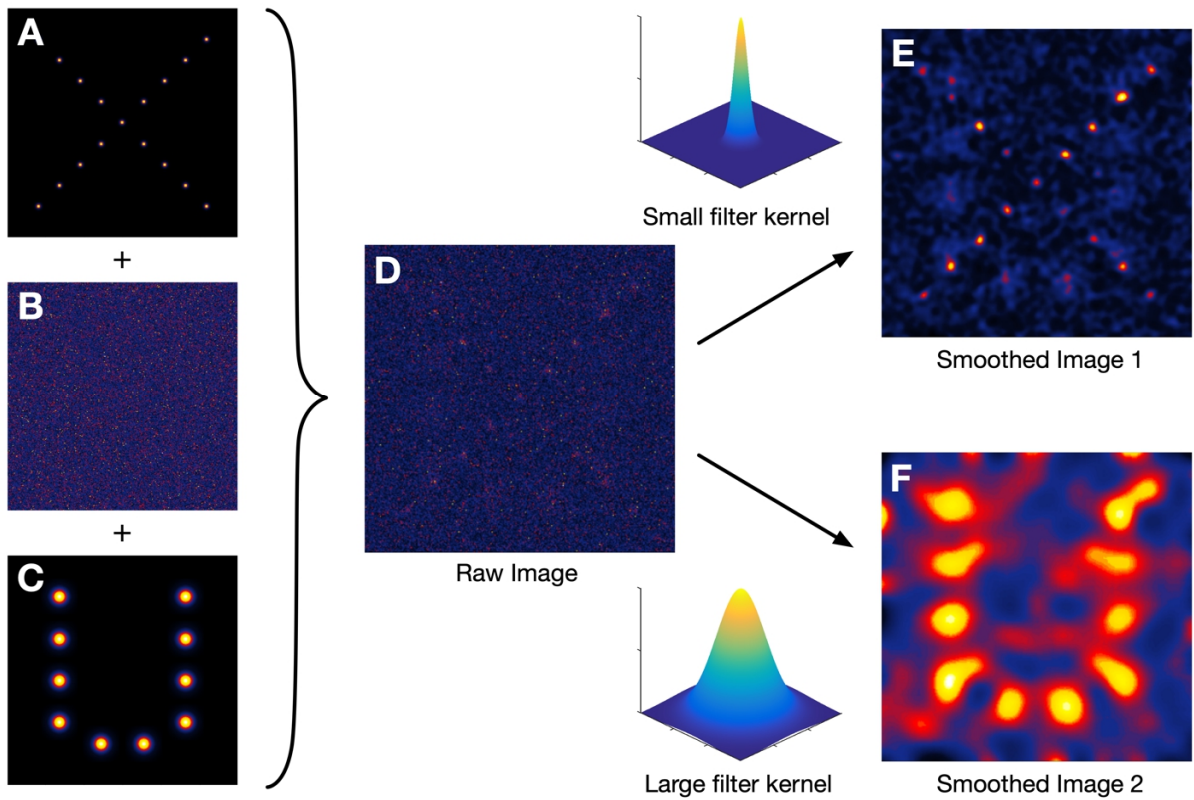

**Fig. 1 Illustration of the matched filter theorem for image analysis using Gaussian smoothing kernels.** When combining images containing a specific pixel pattern ("X" panel **A,** "U" panel **C**) with an image of random pixels (panel **B**) into a raw image (panel **D**) that contains both signal and noise (much like a raw echo-planar image from fMRI that contains a BOLD signal). When applying smoothing kernels of different size (at full width at half maximum; see arrows) as in fMRI preprocessing, the resulting images can only resolve pixel patterns at the respective spatial observation scale (panels **E** and **F**). Figure inspired by Ball et al., 2012 (8).

However, we do not *a priori* know the optimal spatial observation scale for tracking the shape and the size of the BOLD signal in the human brain in Gaussian scale space. Thus, arbitrarily choosing one particular filter size for the analysis of the data may not provide a full account of the underlying brain responses. At the group level, for example, small filter sizes appear to bias the analysis towards regions with little interindividual anatomical variability, whereas larger filters bias the analysis towards regions with greater variability, potentially lumping together functionally separate regions (10, 12).

Despite these concerns, spatial smoothing with a single filter size is still a routine part of the preprocessing of fMRI data. In a study of methods reporting in a mixed sample of 241 fMRI studies,

Carp *et al.*, 2012 (13) found that 88% of studies reported using spatial smoothing. However, only 8% of those studies provided a rationale for the size of the specific smoothing kernel chosen, and the most common (46.3%) filter size was the 8-mm (full width at half maximum, FWHM) filter (13) provided by default in the "Statistical Parametric Mapping" (SPM) software package (14). Overall, almost all fMRI studies use the default filter size of their analysis software (e.g., 8 mm FWHM), within a narrow range of filters between 2 and 12 mm FWHM (13).

To address this problem of "spatial tunnel vision" in conventional monoscale fMRI data analysis, spatial scale-space searches through the application of Gaussian filters of different sizes, have been proposed as possible solutions (15, 16). However, this method has very rarely been used, possibly because of the paucity of studies illustrating their advantages and the computational effort involved. Previous studies involving motor tasks (10, 17) and visuomotor and auditory processing (8, 18) with scale-space searches have demonstrated that there is substantial variability in the anatomical assignment of activation peaks, depending on the filter size. This suggests that multiscale analyses may facilitate a more comprehensive understanding of the full shape and size of the BOLD signal in Gaussian scale-space. Speech-related patterns of fMRI responses—an important application domain of fMRI in cognitive neuroscience—have not yet been investigated in this context.

Here, we use limited spatial multiscale analyses in SPM across a range of filters from 1 to 20 mm FWHM in a speech repetition fMRI experiment conducted in 25 subjects to demonstrate the crucial importance of multiscale fMRI analyses for neurolinguistics research.

We show that depending on the filter size, the results and the functional interpretation supported by the observed response patterns vary to an astonishing degree. For example, we pinpoint how small spatial filter kernels bias the results towards subcortical and cerebellar activation clusters. We also show how the atlas-based anatomical assignment of suprathreshold peaks in fMRI shifts in relation to the filter width and how this may lead to fundamentally different functional interpretations of the data. We demonstrate how spatial multiscale analyses are required to fully explore the deep structure of the BOLD signal across Gaussian scale-space to better understand large-scale brain networks that support speech processing. Beyond this specific context of speech processing, we advocate for considering the spatial observation scale as a new dimension in the analysis and interpretation of fMRI (and PET) data.

## Results

We first describe the topographical changes of the suprathreshold activity maps as analyzed with the widely used software Statistical Parametric Mapping (SPM) from individual and group analyses across a range of Gaussian filter kernels (1–20 mm FWHM). To illustrate the effects of different filter sizes, we chose the simple contrast "repeating pseudowords" (RPW) > "repeating words" (RW) and hypothesized that repeating unfamiliar pseudowords would reliably engage the perisylvian language processing network. Although more sophisticated methods for analysis, such as parametric modulations or multivariate pattern analyses, are certainly available in SPM, this so-called "classical inference" is still widely used in fMRI research today. We then show how variations in the spatial observation scale affect the number and anatomical assignment of suprathreshold peaks.

*Individual variability of SPM analyses across filter sizes*

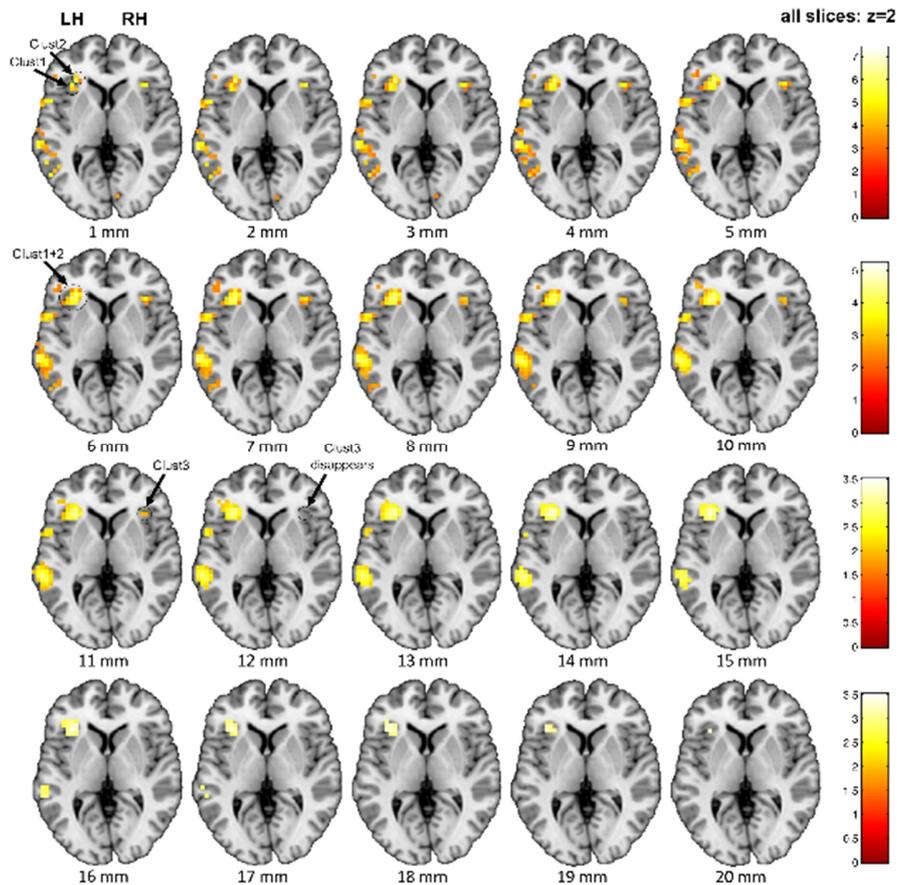

**Fig. 2 Results from individual SPM analyses in one subject (Subject #2 of 25)** across Gaussian smoothing filters, ranging from 1–20 mm (FWHM). T-statistic at p<0.001, uncorrected. Note the different scaling of the color bars. Abbr. Clust=Cluster.

In **Fig. 2**, we see how individual SPM maps change across filters sizes. The plots illustrate several (expected) effects of analyzing the data at different filter sizes:

(a) Scattered small suprathreshold clusters that are visible at small filter sizes merge into larger clusters with increasing filter sizes. Note, for example, how the many small clusters at small filter sizes in the left inferior frontal cortex (**Fig. 2** clusters 1 and 2, indicated with arrows) merge into one, seemingly homogenous, cluster at larger filter sizes.

(b) Small clusters that are visible at smaller filters sizes may completely disappear at larger filter sizes. Note, for example, how the analyses at filter sizes from 1–11 mm contain right hemispheric clusters in the right insula, which then disappear at filter sizes > 11 mm (**Fig. 2** cluster 3, indicated with arrows).

(c) The average t-value at the voxel-level decreases with increasing filter size. We take the t-value here as an illustrative proxy for the "strength" with which the null-hypothesis (i.e., no difference in BOLD signal between RPW and RW) is rejected in each (suprathreshold) voxel.

Moving from the individual to the group level, we show the results from the SPM group analysis in the same contrast (RPW>RW).

*Dynamic of SPM group analyses across filter sizes*

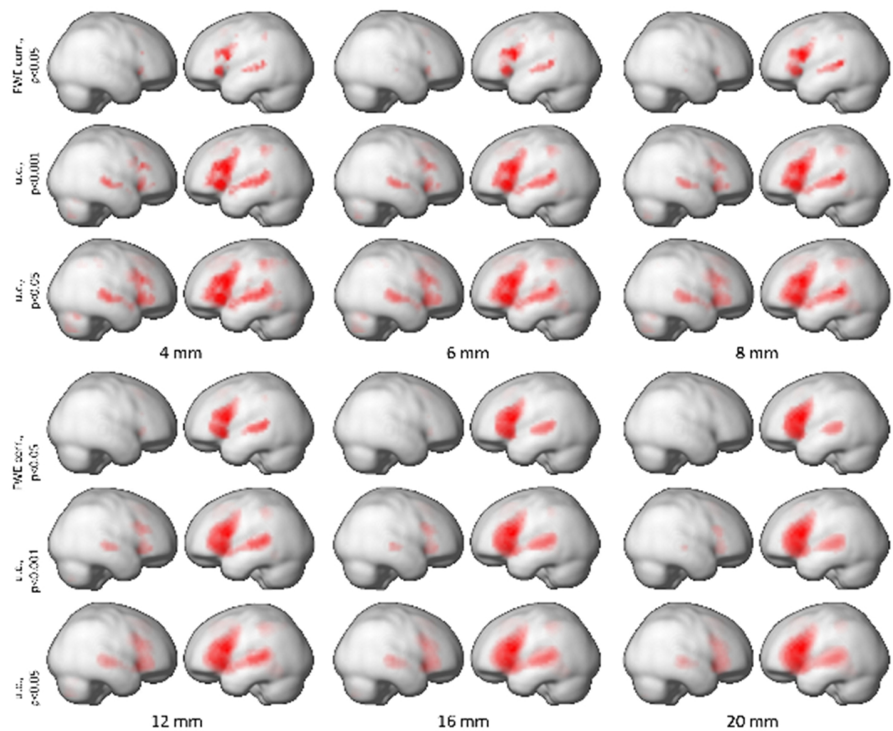

**Fig. 3** T-Maps from fMRI group analyses (n=25 subjects) contrasting the repetition of pseudo-words with the repetition of words. T-statistic at three levels of significance: FWE corrected (p<0.05), uncorrected (u.c.) at p<0.001, and u.c. at p<0.05 across various filter sizes from 4 mm to 20 mm (FWHM).

As with the analyses at the subject level, the group analyses show several effects of filter size across subjects (and at different levels of significance threshold, from liberal to strict):

(a) Small and scattered clusters at small filter sizes merge into larger clusters with increasing filter size.

(b) Initially, small clusters, e.g., in the right hemisphere and cerebellum, disappear depending on the threshold of the significance level at larger filter sizes. For example, while we find suprathreshold clusters in the right ("linguistic") cerebellum at small filter sizes (1–12 mm), the cerebellar clusters do not reach suprathreshold strength (even at p<0.05 uncorrected) at filter sizes above 12 mm.

*Overall variability of cluster size and BOLD signal strength across subjects and filters*

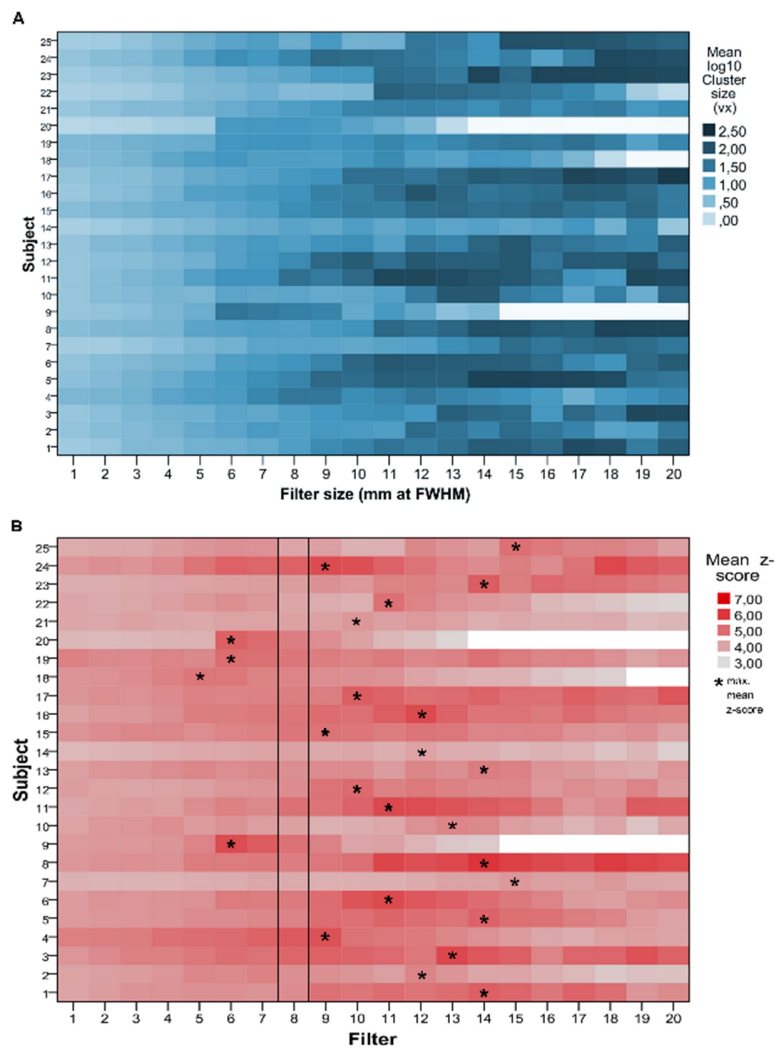

**Fig. 4 Varying the spatial filter size: effects on cluster size and BOLD signal strength. A** shows the variability of the average size of the cortical suprathreshold clusters across subjects (n=25) and filters (1–20 mm FWHM). **B** shows the overall variability of the average BOLD signal strength (mean z-score of all cortical suprathreshold clusters) across subjects and filters. The results are derived from SPM analyses of the contrast of repeating pseudowords > words at the significance level of p<0.001, uncorrected. Notably, we find that in this sample of subjects, at the SPM8 default filter of 8 mm, none of the subjects reached their maximum average BOLD signal strength (indicated by an **\***).

When plotting the overall variability of the average cortical cluster size (mean log$_{10}$ of the number of voxels) across subjects (n=25) and filters (1–20 mm FWHM), we found that cluster size increased consistently in all subjects with increasing filter size—an expected finding, as smaller clusters merge into larger clusters (**Fig. 4A**). The plot reveals, however, that there is substantial variability in cluster size changes across filters. While some subjects (such as #3 or #18) show a steady increase, eventually reaching a maximum mean cluster size at a filter size of 20 mm, in other subjects, such as #10 or #4, mean cluster size fluctuates substantially across filters.

The overall variability of average BOLD signal strength (mean z-score) of cortical suprathreshold clusters across subjects (n=25) and filters (1–20 mm FWHM) is shown in **Fig. 4B**. With few exceptions (subjects #9, #18, #19, #20), most subjects reached the maximum BOLD signal strength at filters in the range of 9 to 15 mm. Notably, we find that at the most widely used SPM default filter of 8 mm (FWHM), *none* of the subjects reached their maximum average BOLD signal strength.

*Variability of cluster size and signal strength of left and right perisylvian regions across filters*

Numerous studies have shown that speech processing involves both left and right perisylvian brain regions (19). Models based on evidence from neuroimaging experiments, however, differ with respect to which specific aspects are computed in which hemisphere, to what degree, and in which context. Therefore, we also investigated whether there were any salient differences in cluster sizes or effect strengths between left and right perisylvian regions with respect to varying filter sizes. To this end, we plotted the average cluster size (**Fig. 4**) and average BOLD signal strength (**Fig. 5**) across subjects in perisylvian areas involved in speech processing.

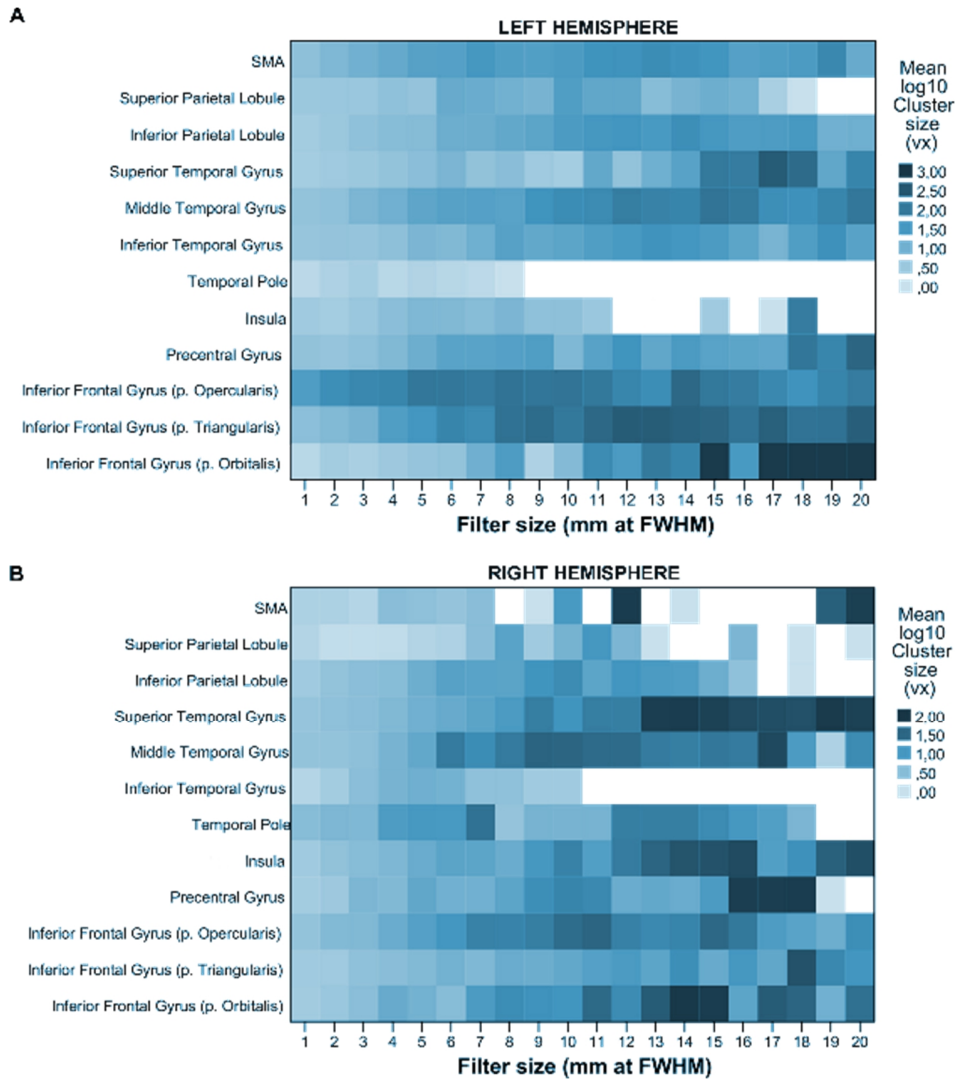

**Fig. 5 Filter size and response cluster size in left and right hemispheric cortical regions**. **A** and **B** show the changes in mean cluster size ($\log_{10}$ voxels) in the subjects (n=25) across spatial filters ranging from 1–20 mm FWHM in left (**A**) and right (**B**) perisylvian regions. The results are derived from SPM analyses of the contrast of repeating pseudowords > words at the significance level of p<0.001, uncorrected.

In **Fig. 5**, we plotted the changes in mean cluster size across filter sizes. As in the analysis of the overall cluster size changes in relation to different filters, we generally found an increase in mean cluster size with increasing filter size. If we look at the specific changes in homotopic left and right regions, however, we see how salient differences between the hemispheres emerge. While the left inferior frontal gyrus, pars opercularis (IFGpo), reaches its maximum (mean) cluster size at a filter of 14 mm, for instance, the right IFGpo reaches a maximum cluster size at 10 mm. To put it succinctly, none of the homotopic regions in this experiment reached the maximum cluster size at the same respective filter size, which indicates that consequences for investigations of cerebral lateralization may potentially be profound if data analysis is restricted to a spatial monoscale with a single filter width.

Next, we analyzed the influence of varying the spatial observation scale on the mean BOLD signal strength in left and right hemispheric perisylvian regions.

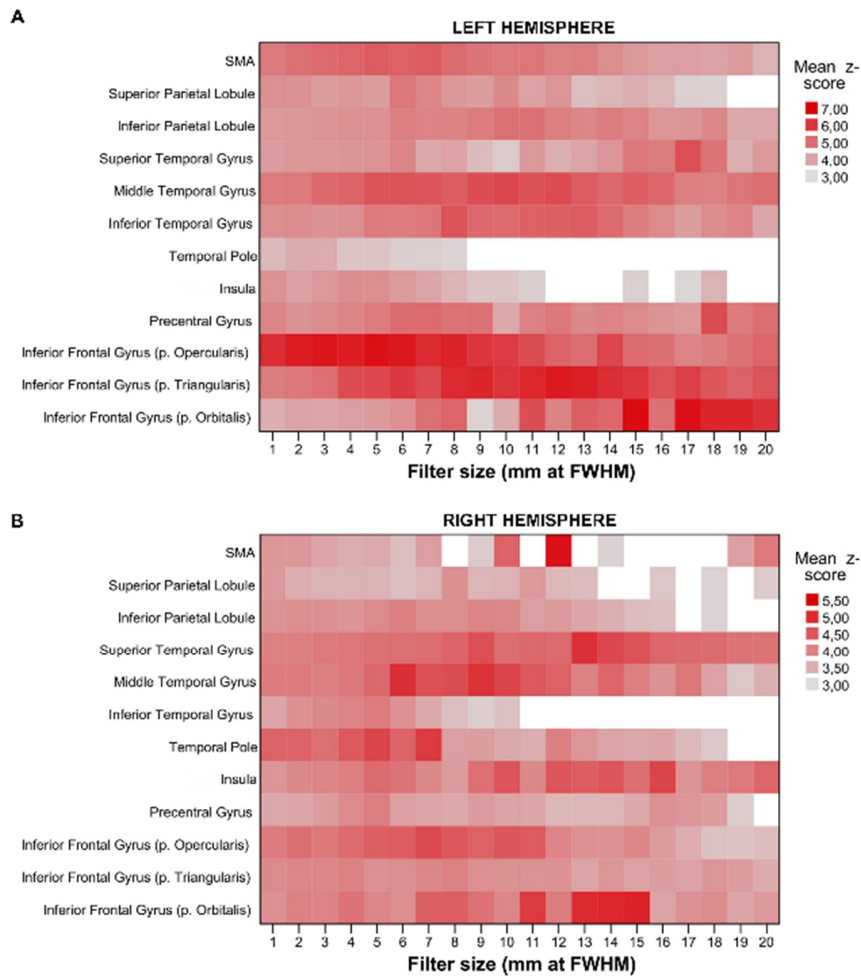

**Fig. 6 Variability of BOLD signal strength in left and right hemispheric cortical regions**. Panels **A** and **B** show the dynamics of mean BOLD signal strength (mean z-score) of the subjects (n=25) across spatial filters ranging from 1–20 mm (FWHM) in left (panel **A**) and right (panel **B**) perisylvian regions. The results are derived from SPM analyses of the contrast of repeating pseudowords > words at the significance level of $p<0.001$, uncorrected.

In **Fig. 6**, changes in mean strength of the suprathreshold BOLD signal (represented by mean z-score) across filter sizes are shown for the left (A) and right (B) hemisphere. As in the *intersubject* variability in signal strength (**Fig. 4B**), these findings show that there is also a substantial *intra-* and *interhemispheric* variability in signal strength depending on the filter size. In each hemisphere, different regions have their maximum (mean) suprathreshold signal strength at different filter sizes (intrahemispheric variability); compare, for example, the left IFGpo and left superior temporal gyrus (STG) or IFGpo and STG in the right hemisphere. Between homotopic regions in the left and right hemisphere, we also see differences with respect to the filter size at which we found the maximum mean signal strength across subjects.

*Changes in anatomical assignment of suprathreshold clusters with varying filter sizes*
To investigate potential shifts in the atlas-based anatomical assignment of suprathreshold peaks, we extracted the anatomical label (at the level of macroanatomical subregions, such as the superior temporal gyrus, based on the probability assignment of the SPM anatomy atlas) for each cluster peak coordinate at the group analysis level for each filter size (1–20 mm FWHM).

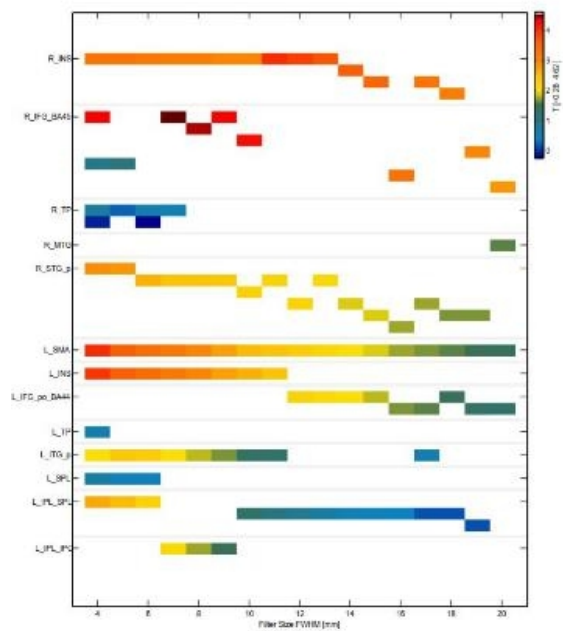

**Fig. 7 Shifts in anatomical assignment with varying filter size.** Mapping of the shifts in anatomical assignment (SPM Anatomy Atlas) across filters (1–20 mm FWHM) from the group analysis of repeating pseudowords > words (n=25 subjects). The t-value (color bar) is included here for a single subject (#2) to exemplify the individual contribution to the group effect size. As an example of an anatomical shift, note how the cluster attributed to the left insula (from filters 1–11 mm) shifts to the left inferior frontal gyrus, BA 44 (filters 12–20 mm).

**Fig. 7** demonstrates how the suprathreshold cluster attributed to the left insula (INS) at filter sizes of 1–11 mm shifts to the adjacent left IFGpo (filter sizes 12–20 mm) and how the assignment of a suprathreshold cluster in the right temporal lobe shifts from the right superior temporal gyrus (STG, filter sizes 1–19 mm) to the right middle temporal gyrus (MTG, filter size 20 mm); these changes in anatomical assignment have profound interpretational consequences, as we will discuss below. Next, we analyzed the changes in the number and proportion of peak coordinates of the suprathreshold clusters depending on the filter size relative to the widely used default filter of 8 mm FWHM in SPM.

*Distributional divergence of peak coordinates across filters relative to the default 8 mm filter*
During fMRI-based connectivity analyses, the number and distribution of the suprathreshold peaks are important parameters, as they will determine the number of nodes and possible connections in any network analysis. Most connectivity analyses, for example Dynamic Causal Modeling (DCM) in SPM, use peaks from suprathreshold clusters from a first-level analysis as seeding nodes for inferring connectivity. Given the spatial-scale-dependent variations we have described here, the filter size may also substantially affect such analyses.

In terms of the distributional divergence of the peak coordinates, we observed the following changes across filter sizes (**Fig. 8A** and **8B**): When looking at the absolute number of peak coordinates, it first appears as if filters smaller than the SPM default of 8 mm (FWHM) result in more spatial divergence of cluster coordinates than filters larger than 8 mm (**Fig. 8A**). This impression changes, however, when we consider the proportion of coordinates per filter size that differ from the default filter at 8 mm. The overall number of activation clusters decreases with filter size as separate smaller clusters merge into larger clusters. Relative to the default 8-mm filter and relative to the absolute number of peaks at each filter, we find that the filter range with the lowest relative spatial divergence lies at filter sizes above 8 mm, mostly from 9 to 13 mm (**Fig. 8B**). Not surprisingly, this filter range between 9 and 13 mm is where most of the subjects' BOLD signal maxima were found (cp. **Fig. 4B**).

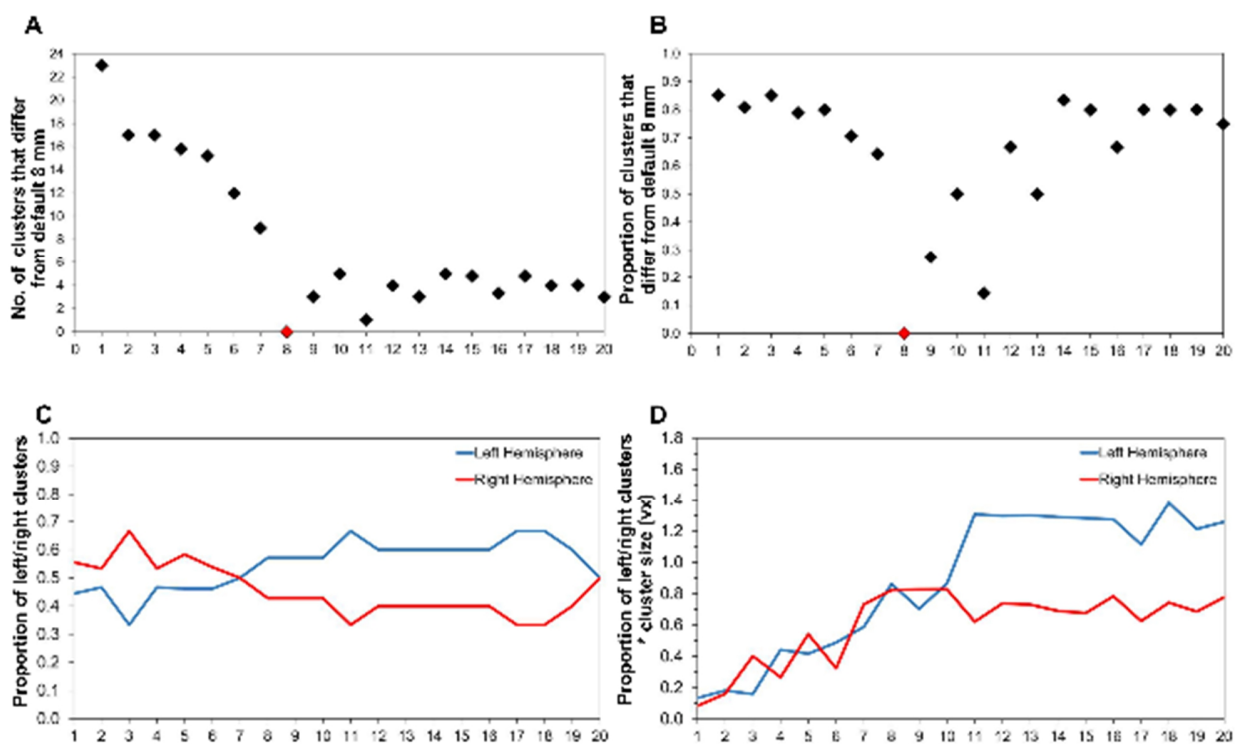

**Fig. 8 Effects of filter size on the number of suprathreshold cortical clusters and left/right distribution.** Panel **A** shows the number of peak voxels (per suprathreshold cluster) per filter size (1–20 mm) that differ in coordinate location (x,y,z in MNI space) from the SPM default filter of 8 mm (red dot). Panel **B** shows the proportion of peak voxels per filter size that differ in their coordinates from the default filter of 8 mm (red dot). Panel **C** shows the proportion of left to right suprathreshold clusters per filter size, and panel **D** shows this proportion corrected for cluster size (voxels). All results are derived from SPM group analyses of the contrast of repeating pseudowords > words at the significance level of p<0.001, uncorrected. The results shown are for clusters in the cerebral cortex only.

Different features of language engage different regions in the language-processing network in the human brain. For cortical regions, speech processing based on phonological, semantic, and syntactic features has a well-established left lateralized bias, whereas the processing of paralinguistic features such as affective prosody or rhythm occurs in right perisylvian cortical regions. More recently, researchers were able to demonstrate that speech processing in both the left and right hemisphere is organized in a dual-pathway architecture (20). Therefore, we also looked at the influence of the spatial filter size on the proportion of left and right suprathreshold clusters in cortical regions as a measure of lateralization in our speech task (**Fig. 8C** and **D**). Computation of the proportion of left/right suprathreshold clusters based on the number of clusters (**Fig. 8C**) would suggest that smaller filters privileged right over left hemispheric clusters, with the expected left lateralized distribution only appearing at filter widths of 8 mm (FWHM) and above. When correcting for mean cluster size (**Fig. 8D**), however, we see an equiproportional distribution of left and right clusters at smaller filter sizes (1–12 mm FWHM) and the emergence of left lateralization only at filter sizes above 12 mm. Importantly, the pattern of changes in average cluster size with increasing filter widths (aside from the consistently larger average cluster size for left over right hemispheric clusters) is essentially the same between left and right hemispheric cortical clusters (**Fig. 9A**). Thus, this emergence of laterality effects at larger filter widths (**Fig. 8D**) is not due to differences in cluster size change with increasing filter width. Instead, we hypothesize that the spatial-scale-dependence might be a more nuanced representation of laterality effects in speech processing than conventional analyses at a spatial monoscale.

*Differences in cluster size across filters between cortical and cerebellar suprathreshold clusters*
In terms of non-cortical regions, many studies have shown that the right "linguistic" cerebellum is also crucially involved in speech processing, for example, for internal forward modeling and the fine tuning of speech movements (21, 22). Therefore, we also investigated the spatial scale-dependence of cerebellar clusters (**Fig. 9B**).

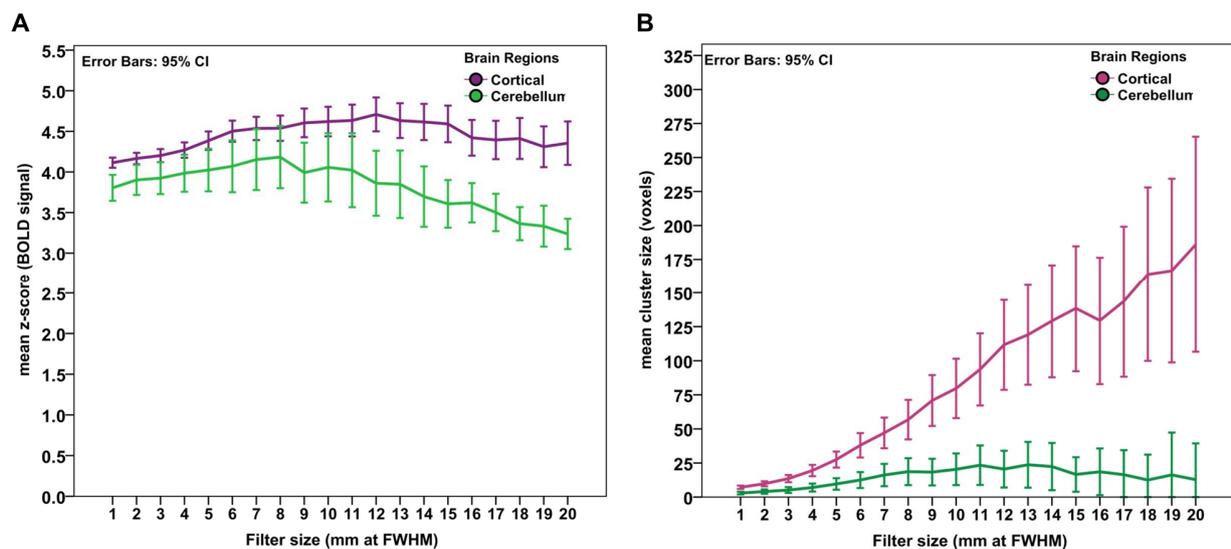

**Fig. 9** Panel **A** shows the average changes in *BOLD signal strength* (represented by mean z-score) of suprathreshold clusters in the cerebral cortex (dark purple) and cerebellum (light green) across filter sizes (1–20 mm FWHM). Panel **B** shows the changes in *cluster size* (in voxels) of suprathreshold clusters in the cerebral cortex (light purple) and cerebellum (dark green) across filter sizes. The results are derived from SPM group analyses (n=25 subjects) of the contrast of repeating pseudowords > words at the significance level of $p<0.001$, uncorrected.

Both the changes in BOLD signal strength (**Fig. 9A**) and cluster size (**Fig. 9B**) from small to large filters showed a surprising dissociation between suprathreshold clusters in the cerebral cortex and the cerebellum. The average BOLD signal strength for clusters in the cerebral cortex remains relatively constant across filters. In the cerebellum, in contrast, the average signal strength remains constant across smaller to medium filters and then decreases from medium to larger filters. For the average cluster size in voxels, the suprathreshold clusters in the cerebral cortex steadily increase with larger filter sizes, while the clusters in the cerebellum did not substantially increase, an observation that we will discuss below.

## Discussion

Here, we show that the analysis of fMRI data of a simple speech repetition task over a range of Gaussian smoothing filter kernels (ranging from 1 to 20 mm FWHM) reveals substantial variability in neuroanatomical localization as well as the size and average signal strength of suprathreshold clusters, depending on the filter size. Taken at face value, each of these different ensembles of regions at any given filter size could be discussed along with (and justified by) the existing literature on speech processing in the human brain. For example, at a filter size of 8-mm FWHM— the most commonly used filter size in fMRI data analysis—suprathreshold clusters in left and right perisylvian regions that corresponded well to aggregated findings from meta-analyses of language-related fMRI studies (19, 23, 24).

Most fMRI studies have only used a single filter for analysis within a narrow range of filter sizes, mostly from 4 to 12 mm FWHM. Considering our present findings, this spatial monoscale analysis appears to limit the interpretation of findings of individual neuroimaging studies and the validity of meta-analyses based on this body of studies.

We first address the spatial scale dimension for the interpretation of fMRI data. We then discuss the problem of anatomical variability and its consequences for modeling and interpreting large-scale brain networks in cognitive and clinical neuroscience. We then suggest considering the spatial observation scale as an additional dimension in fMRI data analysis to improve the inferences from neuroimaging experiments. Finally, we make suggestions on how to further improve fMRI analysis by combining spatial multiscale analysis and cortical-surface-based smoothing.

**The influence of the spatial observation scale on speech-related processing in the brain**

*The anatomical assignment of suprathreshold cluster peaks may change with filter size and substantially influence the interpretation of speech processing in the human brain*

As our analyses show, expanding the filter range may reveal shifts in the neuroanatomical assignment of suprathreshold cluster peaks depending on the filter size. As **Fig. 7** shows, the anatomical localization of suprathreshold clusters may shift substantially depending on the filter size. At a filter size from 1 to 11 mm (FWHM), for instance, we find an (atlas-based) assignment consistently in the left insula in the group analysis contrasting repetition of pseudo-words and words. From the 12 to 20 mm filters, however, this peak shifted to the left inferior gyrus, pars opercularis. To describe the brain network for speech processing, the difference between the involvement of the insula or inferior frontal gyrus pars opercularis (BA44) would trigger very different functional interpretations for the role of these regions in orchestrating speech, both of which can be justified by existing literature on the topic. The role of the insula in speech processing is not agreed upon. Some researchers have claimed an involvement in speech motor programming and articulation (25–27), while others have challenged a specific role of the insular cortex for language and suggest that the insula is a domain-independent hub for error-related processing (28–32). This speech-specific controversy is closely connected to the wider question of whether the insula is a purely a sensory area or whether there exists also an insular motor cortex (33).

The left inferior frontal gyrus, pars opercularis (comprising BA 44 and a small part of BA45), in contrast, is a classical language region—Broca's area—that is associated with mapping phonological information received from the posterior parietal cortex to the inferior frontal speech network (20, 34–37). For closely adjacent regions, such as the insular and inferior frontal cortex or subdivisions of the temporal cortex, suprathreshold clusters that are clearly separable at smaller filters merge with increasing filter size (cp. **Fig. 2** and **Fig. 3**) and the peak coordinate of the suprathreshold shifts within the merged clusters.

Our results show that the question whether the insular cortex is specifically involved in our highly speech-specific cognitive paradigm may depend on the spatial observation scale. Therefore, future studies that investigate domain-specific processes, for example, with the aim to elucidate domain-specific hubs in large-scale networks, should use spatial multiscale analysis.

Given these constraints, how can the scale-dependent variability in neuroanatomical assignment shown here be leveraged to *inform*, rather than merely complicate, the interpretation of fMRI data? A major problem in this context is that for all complex cognitive functions, including speech, memory, motor imagery or others, we do not possess the "ground truth" against which to compare results from any given fMRI experiment. Rather, our view of the functional architecture of cognitive processes is a puzzle composed of many different pieces of data from neuroimaging, neuropsychology, lesion studies, and electrophysiology, to name a few.

*Effects on left and right hemispheric regions involved in speech processing*
The filter size does not only affect local shifts in atlas-based anatomical assignments but also seems to influence the interpretation of hemispheric contributions to networks, in this case, for supporting speech. The right and left inferior frontal cortex (IFC) are both crucial regions for supporting speech processing. For the left inferior frontal cortex (LIFC), aggregated evidence from many studies suggests a distinct function-anatomical organization in which the anatomical tripartition at least roughly, corresponds to a functional processing gradient (35, 38, 39). In this model, the LIFC, pars opercularis/ventral premotor part (po/PMv), is preferentially involved in phonological processing, the overlapping pars opercularis/pars triangularis (po/pt) is involved in syntactic processing, and the ventral pars triangularis (pt) is involved in semantic processing (37, 40). The role of the homotopic right inferior frontal cortex in language processing, however, is much less clear. There is some agreement that the right IFC appears to support the analysis of paralinguistic, so called extrinsic, features of speech such as affective prosody (41–43). To what degree the right perisylvian cortical regions are also involved in computing purely linguistic features of speech and whether they are also organized in a ventral/dorsal network architecture, however, is not clear.

At smaller filter sizes (1–12 mm FWHM), the clusters in IFC are clearly separated into (a) dorsal and posterior (IFG, pars opercularis) and (b) ventral and anterior (IFG, pars triangularis) clusters. These separate clusters correspond well to the prevailing dual-stream model in speech processing of a dorsal stream for mapping phonological information from the temporo-parietal cortex to the IFGpo and a ventral stream for the higher-level modulation of syntactic and semantic aspects of speech (20, 35, 44). Also note that at the group level, the distinction between these clusters is much more clearly delineated in analyses at a strict level of significance (FWE-corrected, $p<0.05$) than at less strict significance levels (**Fig. 3**). Furthermore, the homotopic right hemispheric IFC regions, which are also assumed to be organized in a dual-stream architecture (45), show a different dynamic across filter sizes and significance levels. If we had analyzed the data at the usual single-filter level, say, at the SPM standard 8 mm filter, and reported (and/or based our conclusions upon) the analysis on a strict significance level, the dual-stream architecture of right hemispheric speech processing would have remained unnoticed. Moreover, we could take the results even to argue *against* a dual-stream architecture for differences in processing between pseudowords and words. Such changes across filters may ultimately skew the interpretation of any particular fMRI study (and meta-analyses relying on such monoscale studies) with respect to hemispheric lateralization of specific aspects of language processing—a topic with a long and controversial history in cognitive neuroscience—while multiscale analyses might help resolving at least some of the controversial issues.

*Different filter-size-dependent dynamics of BOLD signal strength and cluster size between the cerebral cortex and the cerebellum*

Another observation from our spatial multiscale analysis of the fMRI data was the difference in BOLD signal strength and cluster size changes between cortical and cerebellar clusters with increasing filter size. As **Fig. 9** shows, the average BOLD signal strength (**Fig. 9A**) remains constant in the cerebral cortex but decreases in the cerebellum with increasing filter size. The cluster size (**Fig. 9B**) steadily increases with larger filters for suprathreshold clusters in the cerebral cortex but not in the cerebellum. Therefore, fMRI language studies that have used a single and comparatively large filter for data analysis and did not report cerebellar activity might have underestimated the involvement of the cerebellum in speech processing.

Crucially, the cerebellum, particularly the right "linguistic" cerebellum, indeed appears to be an important region in the articulatory network for the monitoring and fine tuning of speech movements (46, 47). More specifically, the right cerebellum appears to be involved in matching speech motor output with internal forward models during articulation (21, 22). Considering the lack of a scale-dependent increase in cluster size with increasing filters, data analysis with larger filters could be particularly prone to missing task-related cerebellar activity.

As to why the cerebellar clusters do not increase with cluster size, we consider the following factors:

(a) *Anatomical considerations*: The cerebellar suprathreshold clusters are localized mainly in the areas VIIa (crus I and II) and VIIb. Both areas are part of the "cerebro-cerebellum" and extensively project to the prefrontal and posterior parietal cortex (48). Cytoarchitectonically, the cerebellar cortex has fewer layers (three) compared to the cerebral cortex (six layers) but a similar average thickness (3 mm for the cerebellar (49) and 1-4.5 mm for the cerebral cortex (50)). The smaller filters of 1-5 mm (FWHM) already capture most of the smaller suprathreshold clusters' extent in both cortices. However, given the similar cortical thickness in the cerebellar and cortical cortices, respectively, the cytoarchitecture is not likely to contribute to the different spatial scaling effects. Other anatomical considerations are that the gyration and cortical folding pattern are more compact in the cerebellum when compared to the cortex which might also have an influence on cluster size dynamics in fMRI (51, 52). An important methodological constraint, inherent to the SPM approach to fMRI analysis (but also other analysis approaches), is that the spatial distribution along the cortical surface might be misrepresented because of 3D filtering. This constraint could be resolved, for example, by surface-constrained smoothing or segmentation of the cerebellar cortex for geodesic estimations.

(b) *Metabolic considerations*: The average energy consumption (in total glucose use per minute, μmol/min) is substantially smaller in the cerebellum (ca. 9 μmol/min per 10 billion neurons) than in the cerebral cortex (150 μmol/min per 10 billion neurons) (53). Conceivably, these different metabolic "energy budgets" for the cerebellum and cerebral cortex may influence the strength and/or stability of the BOLD signal across the spatial observation scale as seen in Fig. 9A.

(c) *Number of suprathreshold clusters*: On average, we found only one cluster per filter size in the cerebellar cortex (compared to several clusters in the cerebral cortex). Thus, the merging of smaller clusters into larger clusters might not play a substantial role in the cerebellum as in the cerebral cortex—in which this merging was clearly detectable throughout the increasing filter sizes (although contributions of subthreshold clusters cannot be excluded).

Importantly, all these structural and functional differences between the organization of the cortex and the cerebellum—anatomy, differences in neurovascular coupling—could be scale-sensitive and may lead to differences in analysis outcomes in spatial monoscale analyses. However, the assumption that the highest significance of the signal occurs at a spatially matching filter assumes white noise (for the noise part of the data). Yet, optimal filters do not only dependent on the signal but also the noise and we do not know whether the noise component is also scale-sensitive.

Given the observed differences in cluster size change across filter sizes between the cerebellum and the cerebral cortex, we therefore strongly recommend the routine inclusion of small filters in the data analysis in fMRI to detect cerebellar clusters that might play an important role in particular tasks and subsequent network analyses.

**Summary: Varying the observation scale allows spatial precision analysis of speech-related fMRI data**

As our results demonstrate, varying the spatial observation scale in analyzing the fMRI data yields important information on the functional organization of speech, here, the increased effort in repeating unfamiliar pseudowords in the brain. Analyzing the data with smaller Gaussian spatial filters reveals the involvement of the insula and the right cerebellum in repeating unfamiliar pseudowords when contrasted with repeating regular words—a well-known speech and language specific paradigm. At the conventional scale of spatial observation, however, this involvement of these regions would not have appeared in the analysis suggesting that the conventional scale for spatial filtering (i.e., 8 mm FWHM Gaussian filters) might not be sufficient to resolve the contribution of important hubs in speech and language brain networks. Given this scale-dependent variability in anatomical localization, cluster size dynamics and BOLD signal strength, we may ask what the correct analysis and interpretation of any given set of fMRI data requires.

**How should the spatial-scale-dependent effects on fMRI data analysis be interpreted?**

In the case of the simple speech task here, for example, should we argue that the insula is a crucial hub for orchestrating the increased effort in language processing in repeating pseudo-words, or the inferior frontal gyrus, pars opercularis, or both? How would we justify each interpretation?

First, when analyzing and interpreting fMRI data, the following aspects may improve our "spatial tunnel vision":

(a) The analysis here and in previous studies (8, 10) underscores the importance of analyzing fMRI data on a spatial multiscale in order to obtain a fuller picture of the apparent scale-dependent intra- and interindividual variability of the BOLD signal in fMRI.

(b) Adjusting the spatial observation scale to the region-of-interest (e.g., the cerebellum) may lead to a more comprehensive picture of the range of regions associated with task-related (and resting-state) fMRI paradigms. We therefore advocate for the use of multiple filters in pre-processing fMRI data and an exploration of the full extent of filters used.

Therefore, we suggest that to overcome the apparent variability in scale-dependent signal behavior within brains (across regions) and between brains (across subjects), spatial multiscale analysis is a first helpful step. However, in our view, further improvements and complementary methods for pre-processing and analyzing fMRI data are also necessary. The variability across the spatial observation scale in fMRI further compounds already existing problems in the integration of data on structure-function relationships in the brain to build large-scale models of cognition (and/or brain pathology). Therefore, a granular analysis of fMRI data that explores the available range of spatial filters may be required, especially since the anatomical organization into small-scall, mid-scale and large-scale networks (54) may require different filter sizes for different scales of network analysis.

**The road ahead: Spatial multiscale analysis for brain networks**

Modern approaches to network neuroscience emphasize the importance of a "glocal" perspective, that is, models that capture the patterns of connectivity of both dense local networks and sparse long-range interregional networks for the integration of information (54–56). The functional interpretation of such networks, however, critically depends on the correct identification of regional "hubs" and their anatomical patterns of connectivity (57). Furthermore, different cortical areas appear to possess a unique "connectional fingerprint" (58), an area-specific pattern of cortico-cortical projections. Thus, the difference between one or another region, even different parts of one region, could result in very different network models for any given cognitive function.

Therefore, mapping the full spatial representation of the BOLD signal in Gaussian scale-space would improve the anatomical interpretability of fMRI analyses. To this end, however, a true spatial multiscale analysis would be necessary. While using multiple spatial monoscale analyses, as demonstrated here, may be useful to map the scale-dependent anatomical gross variability in an fMRI experiment, it is not a full multiscale exploration of the BOLD signal in Gaussian scale-space.

For finding the "optimal" filter size to analyze the fMRI data, we suggest to consider the different advantages (and disadvantages) of small, mid-range and larger filters. The smaller filter sizes indeed provide important information on the different kinds of regions involved in speech processing (but may be liable to false positives in very small clusters). Specific limitations of "conventional" fMRI data analysis are, inter alia, that the applicability of large 3D filters is limited because of spillover from other brain parts (cerebrospinal fluid from the ventricular system, white matter). Therefore, to yield useful information from larger filters, surface-constrained filtering ("smoothing") would be a promising strategy. Larger filters may have a yet untapped potential if one can eliminate the increase of noise from other sources of signal.

Intriguingly, the problem of how to fully explore Gaussian scale-space is not unique to neuroimaging research, but occupies other research areas that are reliant on multiresolution image processing and scale-space representation as well, such as cosmology (59, 60) or computer vision research (61–63). For scale-space representation in computer vision, for instance, the resolution of any kernel for spatial smoothing may be used to define a family of derived images (64–66). In principle, two types of scale-space searches can be applied: a "limited" multiscale analysis to compare datasets with different filter widths or a "true" multiscale transformation of a single dataset. The latter approach would allow a coherent, tractable observation of merging or diverging cortical activity patterns from one scale to the next and thus reveal the "deep structure" of the BOLD signal across Gaussian scale-space. One could also use non-Gaussian filters for spatial smoothing, Gaussian scale space, however, is particularly invariant towards shifts in localization and does not create (or enhance) local extrema and has thus become the canonical approach to generating scale-space representations (67, 68, 62). Our limited multiscale analyses support these theoretical assumptions: The "morphology" of suprathreshold clusters and their localization shifts in coordinate space observed across the spatial scale—though potentially highly relevant for the functional interpretation of results—do not appear random or wildly fluctuating.

All these aspects make Gaussian filters well suited to capture the scale-dependent variations in spatial patterns in fMRI data. Yet, such comparisons of progressive scale-space searches have rarely been applied in neuroimaging research. For the implementation of such multiscale analysis in fMRI, it may be necessary to use other kernels, such as wavelets, to satisfy the scale-space axioms and to arrive at a true representation of the BOLD signal across Gaussian scale-space. For the time being, such approaches towards true multiscale analysis remain to be developed (and customized) for the available fMRI software packages (including with methods for controlling for multiple comparisons testing).

Another approach to multiscale representation with a "filter bank" analysis could be using artificial neural networks (with a spatial hierarchy), i.e., purely data-driven representation learning with artificial neural networks for deep learning. For the time being, we encourage neuroimaging researchers to be aware of the multidimensional problem of anatomical variability, to map the intra- and inter-individual spatial-scale-dependent anatomical variability using suitable methods. Moreover, future efforts in neuroimaging methods development should focus on the development of a true multiscale analysis to advance our understanding of the deep structure and dynamics of the BOLD signal in Gaussian scale-space.

## Materials and Methods

### Subjects

Twenty-five healthy subjects participated in the experiment. The study was approved by the local Ethics Committee of the Medical Center, University of Freiburg (Germany), and all subjects provided written informed consent. Subjects were native speakers of German without any history of serious medical, neurological, or psychiatric illnesses (mean age = 24 years, age range = 18–32 years, 13 females). We assessed hand preference with the 10-item version of the Edinburgh Handedness Inventory (69); all subjects had right hand preference.

### fMRI: Stimuli and experimental design

<u>Stimulus preparation</u>: The stimulus set consisted of 96 words (n=48) and pseudo-words (n=48) of one to four syllables (see supplement for full list of stimuli). The words and pseudo-words were matched for phonetic structure and complexity and are based on research on articulatory phonology by the neurolinguist Wolfram Ziegler and colleagues (70). The 48 words and 48 pseudo-words in German were spoken by a voice-trained native female German speaker in a sound-attenuated room and recorded digitally using the open-source software Audacity (http://audacity.sourceforge.net/). Recorded stimulus lengths varied between 500 and 1500 ms. All stimuli were normalized for loudness. We tested for potential differences in the length of the recordings between words and pseudo-words in a one-way analysis of variance (ANOVA) for the factors' lexical status, number of syllables and syllable structure (consonant-vowel, CV; consonant-consonant, CC). We found no significant differences in recording length between words and pseudo-words. As expected, within each lexical category, recording lengths were significantly (n=96, $p<0.001$) longer for bisyllabic stimuli when compared to monosyllabic stimuli, for trisyllabic when compared to bisyllabic stimuli, and for tetrasyllabic stimuli when compared to trisyllabic stimuli. We found no significant differences in the length of the recordings between stimuli with CV and CC structure in pseudo-words or words.

<u>Stimulus presentation</u>: In the scanner, we presented the stimuli binaurally with Presentation software (http://nbs.neuro-bs.com) and MR-compatible headphones. The stimulus order was pseudo-randomized for the factors' lexical status, number of syllables and complexity, and we used three different sets with varying stimulus orders. The presentation sequence of these sets, in turn, was pseudo-randomized across subjects.

We instructed subjects to fixate on a white cross on a dark gray background in the middle of the presentation screen that did not change during the experiment. An interstimulus interval ("jitter") of 500, 1000 or 1500 ms was used (in pseudo-randomized order) to avoid synchronization with the MR pulse and to make the experiment less monotonous and predictable for the subjects. To balance possible speed-accuracy trade-offs, we instructed the subjects to listen attentively and to repeat the stimuli overtly as quickly but also as accurately as possible while minimizing head movement. We also instructed them not to correct themselves during the experiment even in the case of a perceived error in articulation.

**MRI data acquisition**

MRI data were collected on a 3 Tesla Siemens TIM Trio scanner (Siemens, Erlangen, Germany). Each participant completed three sessions with 400 scans per session. We acquired 24 axial slices covering the whole brain in interleaved order with a gradient echo-planar imaging (EPI) T2*-sensitive sequence (voxel size = 4 mm × 4 mm × 4 mm, matrix = 64 × 64 pixel$^2$, TR = 1.5 s, TE = 30 ms, flip angle = 75°). The EPI images were motion-corrected by image realignment with the reference scan using a retrospective 3D algorithm (71).

We also acquired a high-resolution T1 anatomical scan with an MPRAGE sequence (160 slices, voxel size 1 mm × 1 mm × 1 mm matrix = 240 × 240 pixel$^2$, TR = 2.2 s, TE = 2.6 ms). The total scanning time was 47 min.

**fMRI Analysis**

*Pre-processing*

For the pre-processing of the fMRI data, we used the software "Statistical Parametric Mapping", version 8 (SPM8, http://www.fil.ion.ucl.ac.uk/spm/software/spm8/).

We first realigned the EPI images to correct for head movement. We defined excess head movement as >1.5 mm (i.e., one voxel-width); no subject had to be excluded due to excessive head movement. To correct differences in image acquisition time, we then performed slice timing using slice number 12 (of 24) as a reference slice. We then co-registered the first EPI volume of the first session to the high-resolution T1 volume using the "normalized mutual information" function in SPM. We then segmented the co-registered volumes using the high-resolution space template for European participants of the International Consortium for Brain Mapping (ICBM) as implemented in SPM with very light bias regularization (0.0001) and a 60 mm bias FWHM. Next, we normalized the EPIs into the Montreal Neurological Institute (MNI) standard space using the warping parameters estimated in the previous segmentation step. Finally, we created 20 different smoothed datasets using 3D Gaussian kernels ranging from 1 to 20 mm (FWHM) in 1-mm steps with no explicit masking.

*Statistical inferences on the preprocessed fMRI data*

At the single-subject level, we modeled the hemodynamic response to +1 scan after the presentation of the auditory stimuli (words and pseudo-words). We applied a high-pass filter with a cut-off frequency of 1/128 Hz before parameter estimation. For statistical testing, we used a one-sample t-test to assess the hemodynamic response to words and pseudo-words. We computed statistical parametric t-maps at three different levels of significance: $p<0.05$ (uncorrected), $p<0.001$ (uncorrected) and $p<0.05$ (family-wise error [FWE] corrected).

We are aware of the trade-off between limiting false-positives (Type I error) but increasing the likelihood of false-negatives (Type II error) at strict significance levels ($p<0.05$, FWE-corrected) and the high likelihood of false-positives at liberal thresholds ($p<0.001$ and $p<0.05$) (7). However, the three levels of significance we used here are on the spectrum from very liberal to strict statistical thresholding and represent common practice in fMRI analysis. Here, we want to explore whether the effects of varying the spatial filter size are broadly similar across this representative spectrum of statistical thresholding. Therefore, we did not use other more intermediate thresholding techniques that are sometimes used in fMRI analyses, such as thresholding at the cluster level (e.g., allowing only contingent clusters of >10 voxels), combining cluster thresholding with an intermediate significance level (e.g., >10 voxels at $p<0.005$), or using a False-Discovery Rate (FDR) correction (72).

In the results section, we mainly report tests with the threshold of $p<0.001$ (uncorrected). We used the results from the stricter threshold (FEW, $p<0.05$) and less strict threshold (uncorrected, $p<0.05$) to assess whether patterns of scale-related variability are similar at different levels of significance. The "liberal" threshold of $p<0.05$, uncorrected was included specifically to look at trends at very liberal levels (being aware of many false positives in the data). Since the resulting overall patterns of variability were very similar across the three levels of significance, we primarily show the results from the intermediate threshold of p<0.001 (uncorrected).

*Definition of Regions-of-Interest (ROI) and extraction of t-values and z-scores*

We determined the regions of interest (ROI) from the t-maps derived from the group statistic at any given filter size (1–20 mm FWHM). We then extracted the average beta value across all voxels in each ROI at each filter size using the SPM toolbox MarsBaR (http://marsbar.sourceforge.net/). Due to the linguistic nature of the task, we mainly show results from salient perisylvian left and right areas in the frontal, temporal and parietal cortex and the "linguistic" (47) right cerebellum whereas we leave scattered subcortical clusters out of the analysis. We then transformed the values from the t-maps to z-scores, which we used for plotting the results.

*Anatomical assignment of peak coordinates*

An ulterior goal of cognitive neuroscience, whether with neuroimaging methods or in the behavioral neurology of the olden days, is to delineate the functional neuroanatomy of cognition. To relate certain brain regions, histological areas and/or networks to cognitive functions in task-based fMRI, researchers usually rely on computational tools to assign any particular brain coordinate to particular brain regions or areas. These automated tools are usually implemented in software packages such as SPM (which we used here). Therefore, we also investigated whether the shifting filter size would affect the automatic anatomical assignment of the suprathreshold clusters (based on the peak coordinate of each cluster).

To assign the anatomical area to the coordinates of significant peak voxels, we used the probabilistic SPM-based Anatomy Toolbox (version 18) (73). We assigned the peak voxel of each suprathreshold cluster at three different significance levels for each set of smoothed data (from 1–20 mm FWHM). In cases in which a given voxel could be assigned to more than one area within any region (e.g., IFG) with a different probability (e.g., 60% probability for IFG, BA 44 and 30% probability for IFG, BA 45), we chose the area with the higher probability.

**Acknowledgment**

Funding: This work was (partly) supported by the German Ministry of Education and Research (BMBF) grant 13GW0053D (MOTOR-BIC) to the University of Freiburg – Medical Center, and the German Research Foundation (DFG) grant EXC 1086 BrainLinks-BrainTools to the University of Freiburg, Germany.

**References**

1. R. Pauli, *et al.*, Exploring fMRI Results Space: 31 Variants of an fMRI Analysis in AFNI, FSL, and SPM. *Front. Neuroinformatics*, 24 (2016).

2. V. Rajagopalan, E. P. Pioro, Disparate voxel based morphometry (VBM) results between SPM and FSL softwares in ALS patients with frontotemporal dementia: which VBM results to consider? *BMC Neurol.* **15** (2015).

3. K. Kazemi, N. Noorizadeh, Quantitative Comparison of SPM, FSL, and Brainsuite for Brain MR Image Segmentation. *J. Biomed. Phys. Eng.* **4** (2014).

4. A. Eklund, T. E. Nichols, H. Knutsson, Cluster failure: Why fMRI inferences for spatial extent have inflated false-positive rates. *Proc. Natl. Acad. Sci.* **113**, 7900–7905 (2016).

5. A. Eklund, M. Andersson, C. Josephson, M. Johannesson, H. Knutsson, Does parametric fMRI analysis with SPM yield valid results?—An empirical study of 1484 rest datasets. *NeuroImage* **61**, 565–578 (2012).

6. T. D. Wager, M. A. Lindquist, T. E. Nichols, H. Kober, J. X. Van Snellenberg, Evaluating the consistency and specificity of neuroimaging data using meta-analysis. *NeuroImage* **45**, S210–S221 (2009).

7. M. D. Lieberman, W. A. Cunningham, Type I and Type II error concerns in fMRI research: re-balancing the scale. *Soc. Cogn. Affect. Neurosci.*, nsp052 (2009).

8. T. Ball, *et al.*, Variability of fMRI-response patterns at different spatial observation scales. *Hum. Brain Mapp.* **33**, 1155–1171 (2012).

9. A. Andrade, F. Kherif, J. F. Mangin, D. Le Bihan, J. B. Poline, Scale space searches in cortical surface analysis of fMRI data (2001).

10. T. White, *et al.*, Anatomic and Functional Variability: The Effects of Filter Size in Group fMRI Data Analysis. *NeuroImage* **13**, 577–588 (2001).

11. K. J. Worsley, Testing for signals with unknown location and scale in a χ2 random field, with an application to fMRI. *Adv. Appl. Probab.* **33**, 773–793 (2001).

12. F. Crivello, *et al.*, Intersubject Variability in Functional Neuroanatomy of Silent Verb Generation: Assessment by a New Activation Detection Algorithm Based on Amplitude and Size Information. *NeuroImage* **2**, 253–263 (1995).

13. J. Carp, The secret lives of experiments: Methods reporting in the fMRI literature. *NeuroImage* **63**, 289–300 (2012).

14. K. Friston, *et al.*, *SPM8 manual* (Functional Imaging Laboratory, Wellcome Trust Centre for Neuroimaging, Institute of Neurology, University College London, UK, 2013) (May 22, 2013).

15. K. J. Worsley, S. Marrett, P. Neelin, A. C. Evans, Searching scale space for activation in PET images. *Hum. Brain Mapp.* **4**, 74–90 (1996).

16. K. Worsley, M. Wolforth, A. Evans, Scale space searches for a periodic signal in fMRI data with spatially varying hemodynamic response in *Proceedings of BrainMap*, (1997).

17. A. Geissler, *et al.*, Influence of fMRI smoothing procedures on replicability of fine scale motor localization. *NeuroImage* **24**, 323–331 (2005).

18. I. Mutschler, *et al.*, Time Scales of Auditory Habituation in the Amygdala and Cerebral Cortex. *Cereb. Cortex* **20**, 2531–2539 (2010).

19. C. J. Price, The anatomy of language: a review of 100 fMRI studies published in 2009. *Ann. N. Y. Acad. Sci.* **1191**, 62–88 (2010).

20. J. Fridriksson, *et al.*, Revealing the dual streams of speech processing. *Proc. Natl. Acad. Sci.*, 201614038 (2016).

21. T. Ishikawa, S. Tomatsu, J. Izawa, S. Kakei, The cerebro-cerebellum: Could it be loci of forward models? *Neurosci. Res.* **104**, 72–79 (2016).

22. M. Ito, Control of mental activities by internal models in the cerebellum. *Nat. Rev. Neurosci.* **9**, 304–313 (2008).

23. M. Vigneau, *et al.*, Meta-analyzing left hemisphere language areas: Phonology, semantics, and sentence processing. *NeuroImage* **30**, 1414–1432 (2006).

24. M. Vigneau, *et al.*, What is right-hemisphere contribution to phonological, lexico-semantic, and sentence processing?: Insights from a meta-analysis. *NeuroImage* **54**, 577–593 (2011).

25. A. Basilakos, *et al.*, Activity associated with speech articulation measured through direct cortical recordings. *Brain Lang.* **169**, 1–7 (2017).

26. N. F. Dronkers, A new brain region for coordinating speech articulation. *Nature* **384**, 159–161 (1996).

27. J. V. Baldo, D. P. Wilkins, J. Ogar, S. Willock, N. F. Dronkers, Role of the precentral gyrus of the insula in complex articulation. *Cortex* **47**, 800–807 (2011).

28. E. Fedorenko, P. Fillmore, K. Smith, L. Bonilha, J. Fridriksson, The superior precentral gyrus of the insula does not appear to be functionally specialized for articulation. *J. Neurophysiol.* **113**, 2376–2382 (2015).

29. J. Graff-Radford, *et al.*, The neuroanatomy of pure apraxia of speech in stroke. *Brain Lang.* **129**, 43–46 (2014).

30. J. D. Richardson, P. Fillmore, C. Rorden, L. L. LaPointe, J. Fridriksson, Re-establishing Broca's initial findings. *Brain Lang.* **123**, 125–130 (2012).

31. H. Ackermann, A. Riecker, The contribution(s) of the insula to speech production: a review of the clinical and functional imaging literature. *Brain Struct. Funct.* **214**, 419–433 (2010).

32. A. E. Hillis, *et al.*, Re-examining the brain regions crucial for orchestrating speech articulation. *Brain* **127**, 1479–1487 (2004).

33. I. Mutschler, *et al.*, Functional organization of the human anterior insular cortex. *Neurosci. Lett.* **457**, 66–70 (2009).

34. F. Alonso, J. Sweet, J. Miller, Speech mapping using depth electrodes: The "electric Wada." *Clin. Neurol. Neurosurg.* **144**, 88–90 (2016).

35. P. Kellmeyer, *et al.*, Fronto-parietal dorsal and ventral pathways in the context of different linguistic manipulations. *Brain Lang.* **127**, 241–250 (2013).

36. G. Hickok, The cortical organization of speech processing: Feedback control and predictive coding the context of a dual-stream model. *J. Commun. Disord.* **45**, 393–402 (2012).

37. D. B. Shalom, D. Poeppel, Functional Anatomic Models of Language: Assembling the Pieces. *The Neuroscientist* **14**, 119–127 (2008).

38. S. Bookheimer, FUNCTIONAL MRI OF LANGUAGE: New Approaches to Understanding the Cortical Organization of Semantic Processing. *Annu. Rev. Neurosci.* **25**, 151–188 (2002).

39. D. Poeppel, K. Emmorey, G. Hickok, L. Pylkkänen, Towards a new neurobiology of language. *J. Neurosci. Off. J. Soc. Neurosci.* **32**, 14125–14131 (2012).

40. M. Dapretto, S. Y. Bookheimer, Form and Content: Dissociating Syntax and Semantics in Sentence Comprehension. *Neuron* **24**, 427–432 (1999).

41. M. Belyk, S. Brown, Perception of affective and linguistic prosody: an ALE meta-analysis of neuroimaging studies. *Soc. Cogn. Affect. Neurosci.* **9**, 1395–1403 (2014).

42. E. D. Ross, Nonverbal aspects of language. *Neurol. Clin.* **11**, 9–23 (1993).

43. K. M. Heilman, D. Bowers, L. Speedie, H. B. Coslett, Comprehension of affective and nonaffective prosody. *Neurology* **34**, 917–921 (1984).

44. D. Saur, *et al.*, Ventral and dorsal pathways for language. *Proc. Natl. Acad. Sci. U. S. A.* **105**, 18035–18040 (2008).

45. D. Sammler, M.-H. Grosbras, A. Anwander, P. E. G. Bestelmeyer, P. Belin, Dorsal and Ventral Pathways for Prosody. *Curr. Biol.* **25**, 3079–3085 (2015).

46. J. A. Fiez, The cerebellum and language: Persistent themes and findings. *Brain Lang.* **161**, 1–3 (2016).

47. P. L. Strick, R. P. Dum, J. A. Fiez, Cerebellum and Nonmotor Function. *Annu. Rev. Neurosci.* **32**, 413–434 (2009).

48. C. J. Stoodley, J. D. Schmahmann, Evidence for topographic organization in the cerebellum of motor control versus cognitive and affective processing. *Cortex J. Devoted Study Nerv. Syst. Behav.* **46**, 831–844 (2010).

49. V. Braitenberg, R. P. Atwood, Morphological observations on the cerebellar cortex. *J. Comp. Neurol.* **109**, 1–33 (1958).

50. B. Fischl, A. M. Dale, Measuring the thickness of the human cerebral cortex from magnetic resonance images. *Proc. Natl. Acad. Sci. U. S. A.* **97**, 11050–11055 (2000).

51. M. Hibi, T. Shimizu, Development of the cerebellum and cerebellar neural circuits. *Dev. Neurobiol.* **72**, 282–301 (2012).

52. D. C. Van Essen, C. J. Donahue, M. F. Glasser, Development and Evolution of Cerebral and Cerebellar Cortex. *Brain. Behav. Evol.* **91**, 158–169 (2018).

53. S. Herculano-Houzel, Scaling of brain metabolism with a fixed energy budget per neuron: implications for neuronal activity, plasticity and evolution. *PloS One* **6**, e17514 (2011).

54. E. Bullmore, O. Sporns, Complex brain networks: graph theoretical analysis of structural and functional systems. *Nat. Rev. Neurosci.* **10**, 186–198 (2009).

55. B. Mišić, *et al.*, Network-Level Structure-Function Relationships in Human Neocortex. *Cereb. Cortex* **26**, 3285–3296 (2016).

56. J. Faskowitz, R. F. Betzel, O. Sporns, Edges in brain networks: Contributions to models of structure and function. *Netw. Neurosci.*, 1–28 (2021).

57. D. Mears, H. B. Pollard, Network science and the human brain: Using graph theory to understand the brain and one of its hubs, the amygdala, in health and disease. *J. Neurosci. Res.*, n/a-n/a (2016).

58. R. E. Passingham, K. E. Stephan, R. Kötter, The anatomical basis of functional localization in the cortex. *Nat. Rev. Neurosci.* **3**, 606–616 (2002).

59. H. Morgan, M. Druckmüller, Multi-Scale Gaussian Normalization for Solar Image Processing. *Sol. Phys.* **289**, 2945–2955 (2014).

60. A. Jenkins, A new way of setting the phases for cosmological multiscale Gaussian initial conditions. *Mon. Not. R. Astron. Soc.* **434**, 2094–2120 (2013).

61. C. Lopez-Molina, B. De Baets, H. Bustince, J. Sanz, E. Barrenechea, Multiscale edge detection based on Gaussian smoothing and edge tracking. *Knowl.-Based Syst.* **44**, 101–111 (2013).

62. J. Babaud, A. P. Witkin, M. Baudin, R. O. Duda, Uniqueness of the Gaussian Kernel for Scale-Space Filtering. *IEEE Trans. Pattern Anal. Mach. Intell.* **PAMI-8**, 26–33 (1986).

63. T. Lindeberg, *Scale-Space Theory in Computer Vision* (Springer, 1993).

64. T. Iijima, Basic theory of pattern normalization (for the case of a one dimensional pattern). *Bull Electrotechn Lab* **26**, 368–388 (1962).

65. J. J. Koenderink, The structure of images. *Biol. Cybern.* **50**, 363–370 (1984).

66. A. P. Witkin, Scale-space filtering in *Proceedings of the Eighth International Joint Conference on Artificial Intelligence - Volume 2*, (Morgan Kaufmann Publishers Inc., 1983), pp. 1019–1022.

67. A. Kuijper, L. M. J. Florack, The hierarchical structure of images. *IEEE Trans. Image Process.* **12**, 1067–1079 (2003).

68. L. M. J. Florack, B. M. ter Haar Romeny, J. J. Koenderink, M. A. Viergever, Scale and the differential structure of images. *Image Vis. Comput.* **10**, 376–388 (1992).

69. R. C. Oldfield, The assessment and analysis of handedness: the Edinburgh inventory. *Neuropsychologia* **9**, 97–113 (1971).

70. J. Kappes, A. Baumgaertner, C. Peschke, W. Ziegler, Unintended imitation in nonword repetition. *Brain Lang* **111**, 140–51 (2009).

71. S. Thesen, O. Heid, E. Mueller, L. R. Schad, Prospective acquisition correction for head motion with image-based tracking for real-time fMRI. *Magn. Reson. Med.* **44**, 457–465 (2000).

72. C. R. Genovese, N. A. Lazar, T. Nichols, Thresholding of Statistical Maps in Functional Neuroimaging Using the False Discovery Rate. *NeuroImage* **15**, 870–878 (2002).

73. S. B. Eickhoff, *et al.*, A new SPM toolbox for combining probabilistic cytoarchitectonic maps and functional imaging data. *NeuroImage* **25**, 1325–1335 (2005).